\documentclass[prlprintnumbers,showpacs,reprint,prl,twocolumn,floatfix,superscriptaddress,amsmath,amssymb]{revtex4-1}
\usepackage{epsfig}
\usepackage{subfigure}
\usepackage{amsmath}
\usepackage[usenames,dvipsnames]{color}
\usepackage{amssymb}
\usepackage{setspace}
\usepackage{graphicx} 
\usepackage{dcolumn}
\usepackage{bm}
\usepackage{times}
\input epsf
\usepackage{hyperref}
\hypersetup{
    bookmarks=true,         
    unicode=false,          
    pdftoolbar=true,        
    pdfmenubar=true,        
    pdffitwindow=false,     
    pdfstartview={FitH},    
    pdftitle={My title},    
    pdfauthor={Daniel Larremore},     
    pdfsubject={Research Journal},   
    pdfcreator={},   
    pdfproducer={}, 
    pdfkeywords={} {} {}, 
    pdfnewwindow=true,      
    colorlinks=true,       
    linkcolor=red,          
    citecolor=Green,        
    filecolor=magenta,      
    urlcolor=cyan           
}
\begin{document}
\author{Daniel B. Larremore}
\affiliation{Department of Epidemiology, Harvard School of Public Health, Boston, MA 02115, USA}
\affiliation{Center for Communicable Disease Dynamics, Harvard School of Public Health, Boston, MA 02115, USA}

\author{Woodrow L. Shew}
\affiliation{Department of Physics, University of Arkansas, Fayetteville, AR 72701, USA}

\author{Edward Ott}
\affiliation{Institute for Research in Electronics and Applied Physics, University of Maryland, College Park,
MD 20742, USA}

\author{Francesco Sorrentino}
\affiliation{Department of Mechanical Engineering, University of New Mexico, Albuquerque, NM 87106, USA}

\author{Juan G. Restrepo}
\affiliation{Department of Applied Mathematics, University of Colorado, Boulder, CO 80309, USA}

\begin{abstract}
The collective dynamics of a network of excitable nodes changes dramatically when inhibitory nodes are introduced. We consider inhibitory nodes which may be activated just like excitatory nodes but, upon activating, decrease the probability of activation of network neighbors. We show that, although the direct effect of inhibitory nodes is to decrease activity, the collective dynamics becomes self-sustaining. We explain this counterintuitive result by defining and analyzing a ``branching function'' which may be thought of as an activity-dependent branching ratio. The shape of the branching function implies that for a range of global coupling parameters dynamics are self-sustaining. Within the self-sustaining region of parameter space lies a critical line along which dynamics take the form of avalanches with universal scaling of size and duration, embedded in ceaseless timeseries of activity. Our analyses, confirmed by numerical simulation, suggest that inhibition may play a counterintuitive role in excitable networks. 
\end{abstract}

\title{Inhibition causes ceaseless dynamics in networks of excitable nodes}

\pacs{??}

\maketitle

Networks of excitable nodes have been successfully used to model a variety of phenomena, including reaction-diffusion systems \cite{greenberg}, economic trade crises \cite{arenas}, epidemics \cite{karrer,vanmieghem2012}, and social trends \cite{dodds}. They have also been used widely in the physics literature to study and predict neuroscientific phenomena \cite{kinouchi,wu2007,gollo2009,gollo2012,larremore2011a,larremore2011b,larremore2012}, and have been used directly in the neuroscience literature to study the collective dynamics of tissue from the mammalian cortex in humans \cite{poil2008}, monkeys \cite{shew2011}, and rats \cite{ribeiro,shew2009,shew2011,beggs}. The effects of inhibitory nodes, i.e. nodes that suppress activity, can be important but are not well understood in many of these systems. 
In this Letter, we extend such networks of purely excitatory nodes to include inhibitory nodes whose effect, on activation, is to decrease the probability that their network neighbors will become excited. We focus on the regimes near the critical point of a nonequilibrium phase transition that has been of interest in research on optimized dynamic range \cite{kinouchi,larremore2011a,larremore2011b,shew2009,gollo2009,gollo2012,wu2007}, information capacity \cite{shew2011}, and neuronal avalanches \cite{shew2011,shew2009,petermann,beggs,poil2008,ribeiro}, and has also been explored in epidemiology where it constitutes the epidemic threshold \cite{vanmieghem2012}. At first pass, one would expect the inclusion of inhibition in excitable networks to lead to lower overall network activity, yet we find that the opposite is true: the inclusion of inhibitory nodes in our model leads to effectively ceaseless network activity for networks maintained at or near the critical state. 



Our model consists of a sparse network of $N$ excitable nodes. At each discrete time step $t$, each node $m$ may be in one of two states $s_{m}(t)=0$ or $s_{m}(t)=1$, corresponding to quiescent or active respectively. When a node $m$ is in the active state $s_{m}(t)=1$, node $n$ receives an input of strength $A_{nm}$. Each node $m$ is either excitatory or inhibitory, respectively corresponding to $A_{nm} \geq 0$ or $A_{nm} \leq 0$ for all $n$. If there is no connection from node $m$ to node $n$, then $A_{nm} = 0$. Each node $n$ sums its inputs at time $t$ and passes them through a transfer function $\sigma(\cdot)$ so that its state at time $t+1$ is 
\begin{equation}
	s_{n} (t+1) = 1 \text{ with probability } \sigma \left ( \sum_{m=1}^N A_{nm}s_m(t) \right ),
	\label{eq-update}
\end{equation}
and $0$ otherwise,
where the transfer function is piecewise linear; $\sigma(x)=0$ for $x\leq0$, $\sigma(x)=x$ for $0<x<1$, and $\sigma(x)=1$ for $x \geq 1$.  In the presence of net excitatory input, a node may become active, but in the absence of input, or in the presence of net inhibitory input, a node never becomes active.

We consider the dynamics described above on networks drawn from the ensemble of directed random networks, where the probability that each node $m$ connects to each other node $n$ is $p$. In a network of $N$ nodes, this results in a mean in-degree and out-degree of $\langle k \rangle = Np$.  First, to create the matrix $A$, each nonzero connection strength $A_{mn}$ is independently drawn from a distribution of positive numbers.  While our analytical results hold for any distribution with mean $\gamma$, in our simulations the distribution is uniform on $[0,2\gamma]$. Next, a fraction $\alpha$ of the nodes are designated as inhibitory and each column of $A$ that corresponds to the outgoing connections of an inhibitory node is multiplied by $-1$. Many previous studies have shown that dynamics of excitable networks are well-characterized by the largest eigenvalue $\lambda$ of the network adjacency matrix $A$, with criticality occurring at $\lambda=1$ \cite{larremore2011a,larremore2011b,larremore2012,pei2012}. In order to achieve a particular eigenvalue $\lambda$, we use $\gamma = \lambda/\left[\langle k \rangle (1-2\alpha)\right]$, an accurate approximation for large networks \cite{restrepo2007}. We explored a range of $0 \le \alpha \le 0.3$, which includes the fraction $\alpha \approx$~0.2, corresponding to the fraction of inhibitory neurons in mammalian cortex \cite{meinecke}, and note that as $\alpha$ approaches 0.5, $\gamma$ diverges. If excitatory and inhibitory weights are drawn from different distributions, larger fractions $\alpha$ are possible which we discuss in context below Eq.~\eqref{eq-lambda}.

\begin{figure}[t]
	\centering
	\includegraphics[width=0.85\linewidth]{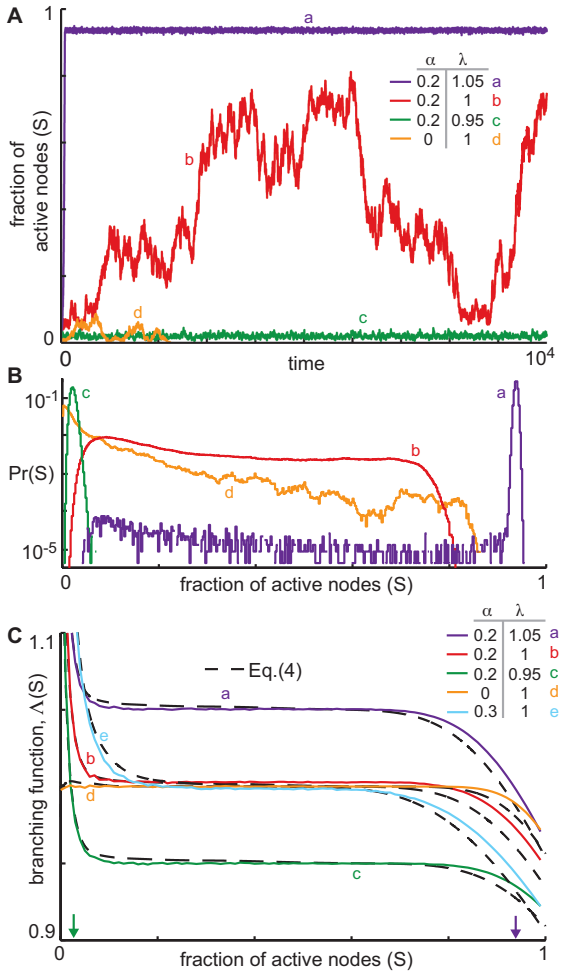}
	\caption{(Color online.) {\bf A)} Time series of $S(t)$ show typical behavior of this system: $\alpha>0$ causes the $S=0$ state to become repelling, so that dynamics are self-sustaining. {\bf B)} Empirical distributions of network activity show that states of critical systems are much more uniformly distributed while sub- and supercritical states fluctuate within tight bands. {\bf C)} Predictions of branching function $\Lambda$ [Eq.~\eqref{eq-lambda}] agree well with empirical measurements of $S(t+1)/S(t)$ for various $\lambda$ and $\alpha$. Three regimes corresponding to $\Lambda > 1$, $\Lambda = 1$ and $\Lambda < 1$ are visible, explaining dynamics from panels A and B. The $\Lambda>1$ regime causes self-sustained behavior. Sub- and supercritical networks achieve $\Lambda=1$ at a single $S$ (arrows), around which dynamics fluctuates tightly; critical networks achieve $\Lambda \approx 1$ over a wide range in $S$, allowing broad fluctuations. $\Lambda<1$ for large values of $S$ preventing activity from completely saturating. $N=10^{4}$, $\langle k \rangle = 200$ for all panels.}
	\label{fig1}
\end{figure}

Our study focuses on the aggregate activity of the network, defined as $S(t) = N^{-1}\sum_{n} s_{n}(t)$, the fraction of nodes that are excited at time $t$. According to Eq.~\eqref{eq-update}, if the entire network is quiescent, $S=0$, it will remain quiescent indefinitely. In the excitatory-only case, the stability of this fixed point has been thoroughly investigated, finding stability for $\lambda \leq 1$ and instability for $\lambda >1$. Many studies have examined this phase transition in activity $S$, finding that many of the interesting properties occur at the critical point $\lambda=1$ such as peak dynamic range \cite{kinouchi,larremore2011a,larremore2011b,pei2012,shew2009} and entropy \cite{shew2011}, and critical avalanches \cite{larremore2012,shew2009,shew2011}, and so our investigation is restricted to values of $\lambda$ near 1.

The main result in this Letter is that when inhibitory nodes are included, the state $S=0$ is unstable. The representative time series of $S(t)$ in Fig.~\ref{fig1}A show that when $\alpha>0$, activity no longer ceases. Subcritical network activity fluctuates within a tight band near $S=0$, supercritical network activity fluctuates within a tight band near $S=1$, and critical network activity fluctuates widely, yet is repelled away from $S=0$. Empirical distributions of system states are shown for each of these cases in Fig.~\ref{fig1}B, highlighting the broad distribution for $\lambda=1$, and narrow distributions otherwise. Importantly, Fig.~\ref{fig1}B also demonstrates that for $\alpha>0$, network activity never reaches $S=0$, while for $\alpha=0$ and $\lambda \leq 1$, activity always eventually dies. A raster plot of self-sustained activity with $\lambda=1$ is provided in Fig.~S2 \cite{supplement}. 

\begin{figure*}[t]
	\centering
	\includegraphics[width=0.85\linewidth]{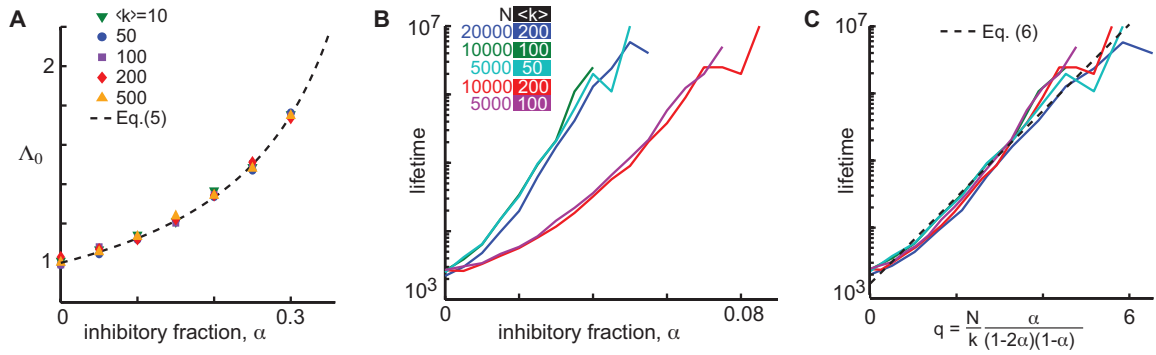}
	\caption{(Color online.) {\bf A)} Empirical measurements of $\Lambda_{0}$ (symbols) agree well with predictions, Eq.~(S15), showing that as $\alpha$ increases, the $S=0$ state becomes more repulsive. {\bf B)} Lifetime of network activity increases with inhibitory fraction $\alpha$ for various $N$ and $\langle k \rangle$. Simulations began with 100 active nodes, with lifetime calculated from the fraction of simulations that ceased prior to $T=10^4$ timesteps. (C) Lifetime scales correctly with $q$, as shown in Eq.~\eqref{eq-scaling}, indicated by collapse of curves.}
	\label{fig2}
\end{figure*}

In order to analyze and understand this behavior, we introduce the {\it branching function} $\Lambda(S)$, which we define as the expected value of $S(t+1)/S(t)$ conditioned on the level of activity $S(t)$ at time $t$, 
\begin{equation}
  \Lambda(S) = S^{-1}\text{E}[S(t+1) \vert S(t) = S].
  \label{lambdadefinition}
\end{equation}
We note that $\Lambda$ is similar to the branching ratio in branching processes except that $\Lambda$ varies with $S$.  For values of $S$ such that $\Lambda(S) >  1$, activity will increase on average, and for values of $S$ such that $\Lambda(S) < 1$, activity will decrease on average. The expectation in Eq.~\eqref{lambdadefinition} is taken over many realizations of the stochastic dynamics. Noting that there is a set of many different possible configurations $\vec{s} = \{ s_n \}_{n=1}^{N}$ of active nodes that result in the same active fraction $S$, we define this set as $\mathcal{S}(S)$. Thus, $\Lambda(S) = S^{-1} \text{E}_{\mathcal{S}(S)} \left [ \text{E} \left [ S(t+1) \vert \vec{s}(t) \in \mathcal{S}(S) \right ] \right ]$, where the outer expectation averages over configurations in $\mathcal{S}(S)$ and the inner expectation averages over realizations of the dynamics for a given configuration. Using Eq.~\eqref{eq-update} we write
\begin{equation}
  \Lambda(S) = S^{-1} \text{E}_{\mathcal{S}(S)} \left [ \left \langle \sigma \bigg ( \sum_{m} A_{nm}s_m(t) \bigg ) \vert \vec{s}(t) \in \mathcal{S}(S) \right \rangle \right ],
  \label{eq-oneexpectationremaining}
\end{equation}
where $\langle \cdot \rangle$ denotes an average over all nodes $n$. $A$ is a large network with uniformly random structure, so we approximate the expectation over $\mathcal{S}(S)$ by assuming each $s_{n}(t)$ is $1$ with probability $S$ and $0$ otherwise, independent of the other nodes. Since nodes differ in the number and type of inputs, this assumption is valid only for large, homogeneous networks. Thus, each node will have, on average, $S \langle k \rangle (1-\alpha)$ active excitatory inputs and $S \langle k \rangle \alpha$ active inhibitory inputs. To account for the variability in the number of such inputs for any particular node (due to both the degree distribution of a random network and the stochasticity of the process),  letting $\mathcal{P}(\beta)$ be a Poisson random variable with mean $\beta$, we model the number of active excitatory inputs as $n_e = \mathcal{P}(S \langle k \rangle (1-\alpha))$ and the number of active inhibitory inputs as $n_i = \mathcal{P}(S \langle k \rangle \alpha)$. We describe the total input to the transfer function using $n_e$ and $n_i$ draws from the link weight distribution.  Replacing the argument of $\sigma$ in Eq.~\eqref{eq-oneexpectationremaining}, and taking the expectation over the distributions of $n_e$ and $n_i$, as well as over the link weight distributions, we approximate
\begin{equation}
  \Lambda(S) \approx S^{-1} \text{E} \left [ \sigma \bigg ( \sum_{j=1}^{n_{e}} w_{j}- \sum_{k=1}^{n_{i}} w_{k} \bigg )  \right ],
  \label{eq-lambda}
\end{equation}
where $w_{j}$ and $w_{k}$ are independent draws from the link weight distribution. Eq~\eqref{eq-lambda} may be used for any function $0 \leq \sigma \leq 1$, and $w_j$ and $w_k$ may represent draws from different excitatory and inhibitory link weight distributions. 

Ceaseless dynamics are now explained by the shape of the branching function, shown in Fig.~\ref{fig1}C. Specifically, for small $S$, $\Lambda(S) > 1$, so low activity levels tend to grow, thus preventing the dynamics from ceasing. The role of inhibition in this growth of low activity may be succinctly quantified as
\begin{equation} 
\Lambda_{0} = \lim_{S \to 0^{+}}\Lambda(S) \approx \lambda\frac{1-\alpha}{1-2 \alpha},
\end{equation}
shown in Fig.~\ref{fig2}A and derived in \cite{supplement}. This estimate coincides with the dominant eigenvalue of the network adjacency matrix without inhibitory links, $\lambda^{+}$, derived in \cite{supplement}. Pei {\it et al.} proposed a different model in which a single inhibitory input is sufficient to suppress all other excitation and found that $\lambda^{+}$ controlled dynamics for all activity levels in their model \cite{pei2012}. In contrast, we find that for moderate values of $S$, $\Lambda(S)\approx\lambda$, and for large values of $S$, $\Lambda(S)$ decreases further. For noncritical networks, $\Lambda(S)=1$ at a single value of $S$, provided $\alpha > (1-\lambda)/(2-\lambda)$. Since $\Lambda(S)$ is non-increasing, $S(t)$ will stochastically fluctuate around that single point of intersection, Fig.~\ref{fig1}C (arrows). On the other hand, for networks in which $\lambda=1$, $\Lambda(S)\approx1$ over a wide domain in $S$, placing the network in a critical state where activity tends to, on average, replicate itself. For large values of $S$, $\Lambda(S) < 1$, imposed by system size.

We find that when there are no inhibitory nodes ($\alpha=0$) network activity resulting from an initial stimulus ceases after a typically short time, in agreement with previous results \cite{kinouchi,larremore2011a,larremore2011b}. However, as $\alpha$ is increased, activity lifetime grows rapidly. To understand the dependence of activity lifetime on model parameters, we simulated the critical case $\lambda=1$ with various $N$, $\langle k \rangle$, and $\alpha$, finding that the expected lifetime of activity after an initial excitation of 100 nodes grows approximately exponentially with increasing $\alpha$, with growth rate proportional to $N/\langle k \rangle$ (Fig.~\ref{fig2}B). Thus large, sparse networks are likely to generate effectively ceaseless activity without any external source of excitation.  The expected lifetime of activity $\tau$, derived analytically (see \cite{supplement}) by treating $S(t)$as undergoing a random walk with drift $(\Lambda(S)-1)S$, is approximately given by 
\begin{equation}
	\tau \sim C_{1} \exp\left\{C_{2} \frac{N}{\langle k \rangle}\frac{\alpha}{(1-2\alpha)(1-\alpha)} \right\},
	\label{eq-scaling}
\end{equation}
where $C_{1}$ and $C_{2}$ are two constants. Figure~\ref{fig2}C shows collapse of numerically estimated $\tau$ for different values of $N/\langle k \rangle$ when plotted against $q~=~N \alpha / \left [ \langle k \rangle (1-2\alpha) (1-\alpha) \right ]$, in agreement with Eq.~\eqref{eq-scaling}.

\begin{figure}[t]
	\centering
	\includegraphics[width=0.85\linewidth]{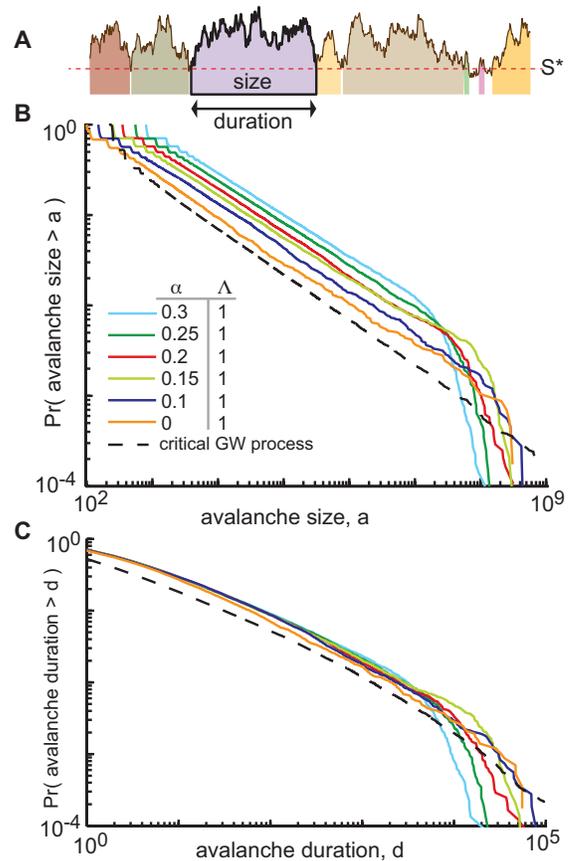}
	\caption{(Color online.) {\bf A)} We define avalanches as excursions above a threshold $S^{*}$, with duration $d$ the length of the excursion and size $a$ the integral under the curve over the duration of the excursion. {\bf B)} Distributions of avalanche size are power law for all $\alpha$, $P(a) \sim a^{-3/2}$. The dashed line corresponds to sizes from a critical Galton-Watson branching process with $S^{*}=128$ . {\bf C)} Durations are not power-law distributed but have the same distribution as durations from a critical Galton-Watson process. Durations do not show the familiar universal power-law exponent of $-2$ due to the conversion of ceaseless time series into avalanches (see text and \cite{supplement}). Data shown: $N=10^4$ nodes over $3\times10^6$ timesteps, $\langle k \rangle = 200$.}
	\label{fig3}
\end{figure}

We now turn our attention to avalanches.  For systems in which activity eventually ceases, an avalanche can be defined as the cascade of activity resulting from an initial stimulus, and thus in excitatory-only models, 
avalanches occur with well-defined beginnings and ends. Because our model generates a single ceaseless cascade, we define an avalanche as an excursion of $S(t)$ above a threshold level $S^*$ \cite{poil2012}, fragmenting a ceaseless timeseries $S(t)$ into many excursions above $S^{*}$, Fig.~\ref{fig3}A. Avalanche duration is defined as the number of time steps $S(t)$ remains above $S^{*}$,  and avalanche size is defined as $a=\sum S(t)$, summing over the duration of the avalanche. This definition corresponds to an intuitive notion of a lower threshold below which instruments fail to accurately resolve a signal. For $\lambda=1$ and all $\alpha$ tested in the model, avalanche sizes are power-law distributed (Fig.~\ref{fig3}B) with exponents that are consistent with critical branching processes and models of critical avalanches in networks \cite{larremore2012}, with size distribution $P(a)\sim a^{-\beta}$ with $\beta \approx 1.5$. This is equivalent to a complementary cumulative distribution function $P($avalanche size $>a)\sim a^{-1/2}$ as displayed in Figure \ref{fig3}B. Exponents from numerical experiments \cite{clauset} are shown in Table S1.  

Critical branching processes \cite{gw} and critical avalanches in excitatory-only networks \cite{larremore2012} should have durations distributed according to a power law with exponent $-2$. However, as can be seen in Fig.~\ref{fig3}C, avalanche durations, while broadly distributed, are not power laws, which we confirmed statistically \cite{clauset}. Though at first glance this appears to disqualify dynamics as critical, we find that time series from a Galton-Watson critical branching process \cite{gw} that are fragmented into avalanches by thresholding show distributions like those shown in Fig.~\ref{fig3}C, and not a power law with exponent $-2$ \cite{supplement}. Our predictions in both Figs.~\ref{fig3}B and C therefore agree well with the criticality hypothesis (dashed lines). Our choice of $S^{*}$ for cascade detection was the lowest value of $S$ for which $\Lambda(S) < 1.01$, thus accounting for differences in the dynamics of the model for different $\alpha$ and acknowledging that for low activity, dynamics are not expected to be critical since $\Lambda(S)$ is far from unity. These results are robust to moderate increases in $S^{*}$.  Based on these observations, we note that to classify or disqualify dynamics as ``critical'' or ``not critical'' based on avalanche duration statistics may depend on precisely how avalanches are defined and measured. 


The inclusion of inhibition in this simple model produces dynamics that may naturally vary between regimes. The low activity regime, where $\Lambda(S) > 1$, prevents activity from ceasing entirely while the high activity regime, where $\Lambda(S)<1$, prevents activity from completely saturating. This may be understood in the following way. For an inhibitory node to affect network dynamics, it must inhibit a node that has also received an excitatory input. When network activity is very low, the probability of receiving a single input is small, and the probability of receiving both an excitatory and an inhibitory input is negligible. Thus, as network activity approaches zero, the effect of inhibition wanes and dynamics are governed by $\lambda^{+}$. On the other hand, when network activity is very high, some nodes receive input in excess of the minimum necessary input to fire with probability one, and so input is ``wasted'' by exciting nodes that would become excited anyway, shifting the excitation-inhibition balance toward inhibition, $\Lambda(S) < 1$. The moderate activity regime, where $\Lambda(S)\approx 1$, features activity that is on average self-replicating. For super- and subcritical networks, the moderate activity regime is a single point, but for critical networks where $\lambda=1$, this regime is stretched, allowing for long fluctuations that emerge as critical avalanches. Thus, for large, critical networks, we find avalanches embedded in self-sustaining activity. 

To conclude, in this Letter we have described and analyzed a system in which the addition of inhibitory nodes leads to ceaseless activity. Our findings may be particularly useful in neuroscience, where self-sustaining critical dynamics has been observed \cite{petermann}. In experiments, networks of neurons exhibit ceaseless dynamics and optimized function (dynamic range and information capacity) under conditions where power-law avalanches occur \cite{shew2009,shew2011,petermann}, but it is not currently possible to directly test the relationship between cortical inhibition and sustained activity {\it in vivo}. One alternative may be to compare empirically measured branching functions from {\it in vivo} recordings with their {\it in vitro} counterparts, where more manipulation of cell populations is possible. This could also be done in model networks of leaky integrate-and-fire neurons, but while criticality \cite{millman} and self-sustained activity without avalanches \cite{vogels} have been found separately, they have not yet been found together. The relation of our mechanism to more traditional ``chaotic balanced'' networks studied in computational neuroscience \cite{vanvreeswijk}, and the ability of balanced networks to decorrelate the output of pairs of neurons under external stimulus \cite{renart} remain open. Outside neuroscience, our results may find application in other networks operating at criticality, such as gene interaction networks \cite{torres-sosa}, the internet \cite{sole}, and epidemics in social networks \cite{dodds,davis}. 

\begin{acknowledgments}
We thank Dietmar Plenz and Shan Yu for significant comments on previous versions of the manuscript. D.B.L. was supported by Award Numbers U54GM088558 and R21GM100207 from the National Institute Of General Medical Sciences. The content is solely the responsibility of the authors and does not necessarily represent the official views of the National Institute Of General Medical Sciences or the National Institutes of Health. E.O. was supported by ARO Grant No. W911NF-12-1-0101.
\end{acknowledgments}


\clearpage
\begin{widetext}

\renewcommand{\thefigure}{S\arabic{figure}}
\renewcommand{\thetable}{S\arabic{table}}
\renewcommand{\theequation}{S\arabic{equation}}

\section*{Inhibition causes ceaseless dynamics in networks of excitable nodes\\
Supplementary Material}

\subsection*{Table S1: avalanche size distribution power-law exponents} 
\begin{table}[h]
\centering
ÊÊÊÊ\begin{tabular}{|l|l|}
\hline
ÊÊÊÊÊÊÊÊDataÊÊÊÊÊÊÊÊÊÊÊÊÊÊÊÊÊÊÊ & Avalanche size distribution exponent $x$, P(size) $\propto$ size$^{-x}$ \\ 
\hline
ÊÊÊÊÊÊÊÊ$\alpha$ = 0.0ÊÊÊÊÊÊÊÊÊÊÊ & 1.50ÊÊÊÊÊÊÊÊÊÊÊÊÊÊ \\ 
ÊÊÊÊÊÊÊÊ0.10ÊÊÊÊÊÊÊÊÊÊÊÊÊÊÊÊÊÊÊ & 1.48ÊÊÊÊÊÊÊÊÊÊÊÊÊÊ \\ 
ÊÊÊÊÊÊÊÊ0.15ÊÊÊÊÊÊÊÊÊÊÊÊÊÊÊÊÊÊÊ & 1.47ÊÊÊÊÊÊÊÊÊÊÊÊÊÊ \\ 
ÊÊÊÊÊÊÊÊ0.20ÊÊÊÊÊÊÊÊÊÊÊÊÊÊÊÊÊÊÊ & 1.48ÊÊÊÊÊÊÊÊÊÊÊÊÊÊ \\ 
ÊÊÊÊÊÊÊÊ0.25ÊÊÊÊÊÊÊÊÊÊÊÊÊÊÊÊÊÊÊ & 1.47ÊÊÊÊÊÊÊÊÊÊÊÊÊÊ \\ 
ÊÊÊÊÊÊÊÊ0.30ÊÊÊÊÊÊÊÊÊÊÊÊÊÊÊÊÊÊÊ & 1.47ÊÊÊÊÊÊÊÊÊÊÊÊÊÊ \\ \hline\end{tabular}
  \caption{Power-law exponents for avalanche size distributions, obtained by simulating Eq.~(1) and collecting avalanches as described in the text,  are very close to $-3/2$ for all $\alpha$ tested. Power-law exponents were calculated using maximum likelihood methods \cite{clauset}. }
  \label{exponenttable}
\end{table}

\subsection*{Derivation of lifetime scaling at criticality}  
The lifetime of critical network activity changes as a function of number of nodes $N$, inhibitory fraction $\alpha$, and  mean degree $\langle k \rangle$, hereafter simply $k$ for notational convenience. Our approach is to derive the functional scaling of activity lifetime $\tau$ by using a Fokker-Planck description, examining the distribution of an ensemble of system states to find those from which cessation of activity is likely.  

We treat network activity $S$ as following a random walk between $S=0$ and $S=1$, with a mean change of $S$ per step, i.e. drift, equal to $\left(\Lambda(S)-1\right)S$. To find the diffusion coefficient we imagine that at $t=t_{0}-1$ all systems in the ensemble have the same $S(t_{0}-1)$ and we ask what the ensemble variance $\langle (S-\langle S \rangle )^{2}\rangle$ is at time $t_{0}$, where $\langle \cdot \rangle$ denotes an ensemble average and $S = N^{-1}\sum_{i} s_i$. Assuming each $s_i$ is independently 1 with probability $S$ and 0 otherwise, we get 
\begin{align}
  2 D = \left \langle \left (S - \langle S \rangle \right )^{2} \right \rangle & = N^{-2} \left \langle \left ( \sum_i ( s_i - \langle S \rangle) \right)^{2} \right \rangle  \nonumber \\
  & = N^{-2} \sum_{i,j} \left \langle \left (s_i - \langle S \rangle \right ) \left (s_j - \langle S \rangle \right ) \right \rangle  \nonumber \\
  & = N^{-2} \sum_{i,j} \delta_{i,j} \left \langle \left (s_i - \langle S \rangle \right )^2  \right \rangle  \nonumber \\
  & = N^{-2} \sum_{i} \left [\langle s^{2}_{i} \rangle - 2 \langle s_i \rangle \langle S \rangle + \langle S \rangle ^2 \right ] \nonumber \\
 & = N^{-1} \left (\langle S \rangle - \langle S \rangle^2 \right) = \mathcal{O}\left(N^{-1}\right) \ll 1.
\end{align}
where we have used $\langle s_{i}^{2} \rangle = \langle s_{i} \rangle$ since $s_i = 0$ or $1$. 
Since $\left \langle \left (S - \langle S \rangle \right )^{2} \right \rangle = \mathcal{O} \left (N^{-1} \right)$, typically $\left (S - \langle S \rangle \right) \sim N^{-1/2}$, and thus we substitute $\langle S \rangle \approx S$, concluding that 
\begin{equation}
  D \approx S (1-S) / 2N.
\end{equation}

Having calculated the diffusion coefficient, we continue with a Fokker-Planck approach to calculate the flux of probability corresponding to the cessation of network dynamics. To study the cessation of activity, we need to determine the behavior of $\Lambda(S)$ for small values of $S$. To do this, we expand $\Lambda(S)$ to first order in $S$. The Poisson variables $n_{e}$ and $n_{i}$ in Eq.~(4) will contribute to first order behavior for only $(n_{e},n_{i}) \in \left \{ (1,0), (2,0), (1,1) \right \}$. Inserting these cases into Eq.~(4) we get
\begin{equation}
	\Lambda(S) \approx S^{-1} \text{E} \left [ \sigma(w) P(n_{e}=1,n_{i}=0) + \sigma(w_{1}+w_{2}) P(n_{e} = 2, n_{i} = 0)+ \sigma(w_{1} - w_{2}) P(n_{e} = 1,n_{i} = 1) \right ]
\end{equation}
By replacing the Poisson probabilities for these cases, we get 
\begin{equation}
	\Lambda(S) \approx S^{-1} \text{E} \left [ \sigma(w) k S  (1-\alpha) e^{-k S} + \sigma(w_{1} + w_{2}) \frac{1}{2} k ^{2} S^{2}  (1-\alpha)^{2} e^{-k S } + \sigma(w_{1} - w_{2})  k^{2}S^{2} (1-\alpha) \alpha e^{-kS } \right ]
\end{equation}
We then expand each exponential to first order in $S$ and assume $w$ is uniform in $[ 0,2\gamma ]$ so that $\gamma = \text{E}[w] = \left[ k (1-2 \alpha)\right]^{-1}$. Assuming that $k \gg 1$ so that $\text{E}[w] \ll 1$, we have that $\text{E}[\sigma(w)] = \text{E}[w] = \left [k (1-2 \alpha) \right]^{-1}$ and $\text{E}[\sigma(w_{1} + w_{2})] = 2 \text{E}[w]$. By integration, $\text{E}[w_{1} - w_{2}] = \gamma/3 =\left[3  k (1-2 \alpha)\right]^{-1}$. Substituting, we get
\begin{equation}
	\Lambda(S) - 1 \approx \frac{\alpha}{1-2\alpha} \left [ 1 - \frac{2}{3} kS  (1-\alpha) \right ],
	\label{eqs5}
\end{equation}
and we find that for small $S$, $\Lambda(S) - 1=(\Lambda_0 - 1) \left [ 1 - 2kS  (1-\alpha)/3 \right ].$ We assume that, in general, we have $\Lambda(S)-1 = (\Lambda_0 - 1) f\left[ kS(1-\alpha) \right]$, where $f(0)=1$, $f(x)\to 0$ for large $x \gg 1$. In addition, since Eq.~\ref{eqs5} indicates that the initial decrease of $f$ with $S$ scales with $kS(1-\alpha)$, for the purposes of estimating the scaling of lifetime $\tau$ with $k$ and $\alpha$, we tentatively take the function $f(x)$ to be independent of $\alpha$. Furthermore, we presume that $\Lambda(S)$ approaches one far from $S=0$ for $Sk \gg 1$. This is in accord with Fig. \ref{fig-pofssketch} which shows a plateau centered at $S=1/2$ where $\Lambda(S)=1$ and this plateau extends down to small $S$. Thus $f(x)$ is supposed to be one at $x=0$ and to approach zero for large $x$. 

Next, we use a Fokker-Planck description with diffusion coefficient $D(S) = S(1-S)/(2N)$ and a ``velocity'' in $S$ of $v_{S} = \left (\Lambda(S) - 1\right)S = (\Lambda_0 -1) S f\left[(1-\alpha)kS\right]$ where $v_{S}$ is motivated by $\text{E}[S(t+1) - S(t) | S(t)] = \left[\Lambda\left(S(t)\right) - 1\right] S(t)$. The time-independent Fokker-Planck equation is
\begin{equation}
  \frac{\partial}{\partial S} \bigg [ (\Lambda_0 -1) f\left[(1-\alpha)kS\right] S P(S) \bigg] = \frac{\partial}{\partial S} \left [ D(S) \frac{\partial P(S)}{\partial S} \right ].
  \label{eq-fppde}
\end{equation}
Since $\Lambda(S)$ is substantially above one for small $S$ and substantially below one for $S$ very near $S=1$ (see Fig.~3), we anticipate that $P(S)$ will have the overall qualitative form shown in Fig.~\ref{fig-pofssketch}, and we now attempt to estimate $P(S)$ in the region $S\sim1/k$. For $S \ll 1$, $D(S) \sim S/2N$, and Eq.~\eqref{eq-fppde} becomes
\begin{equation}
  (\Lambda_0 -1) f\left[(1-\alpha)kS\right] S P(S) = \frac{S}{2N} \frac{\partial P(S)}{\partial S}.
\end{equation}
Defining $\hat{S} = (1-\alpha)kS$, we have 
\begin{equation}
	\left[ \frac{2N}{k(1-\alpha)} (\Lambda_{0} - 1) \right] f(\hat{S}) P(S) = \frac{\partial P(S)}{\partial \hat{S}}.
\end{equation}
In order to solve this equation, we define $x(\hat{S}) = \int_{\hat{S}}^{1/2} f(\hat{S}')d\hat{S}'$ and let $\hat{P}(x(\hat{S})) = P(S)$ which gives
\begin{equation}
	- \left[ \frac{2N}{k(1-\alpha)} (\Lambda_{0} - 1) \right] \hat{P}(x) = \frac{\partial \hat{P}(x)}{\partial x}
\end{equation}
which yields
\begin{equation}
	P(S) = P_{0} \exp{\left \{- \frac{2N(\Lambda_{0} - 1)}{k(1-\alpha)} x\left[(1-\alpha)kS\right] \right \}}.
\end{equation}
where $P_{0} \sim P(1/2)$ (cf. Fig.~\ref{fig-pofssketch}).
\begin{figure}
	\centering
	\includegraphics[width=0.48\linewidth]{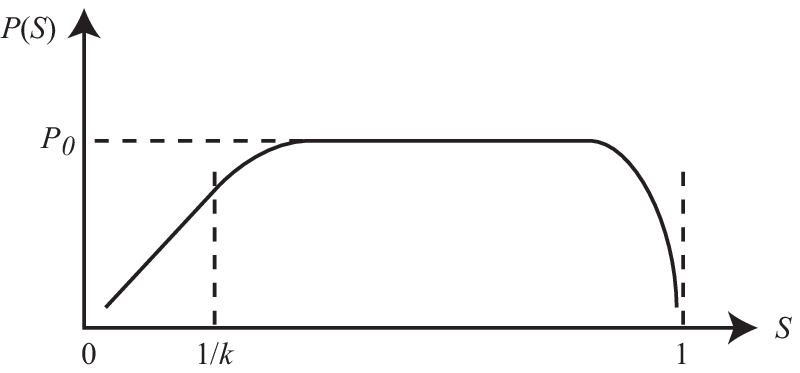}
	\includegraphics[width=0.48\linewidth]{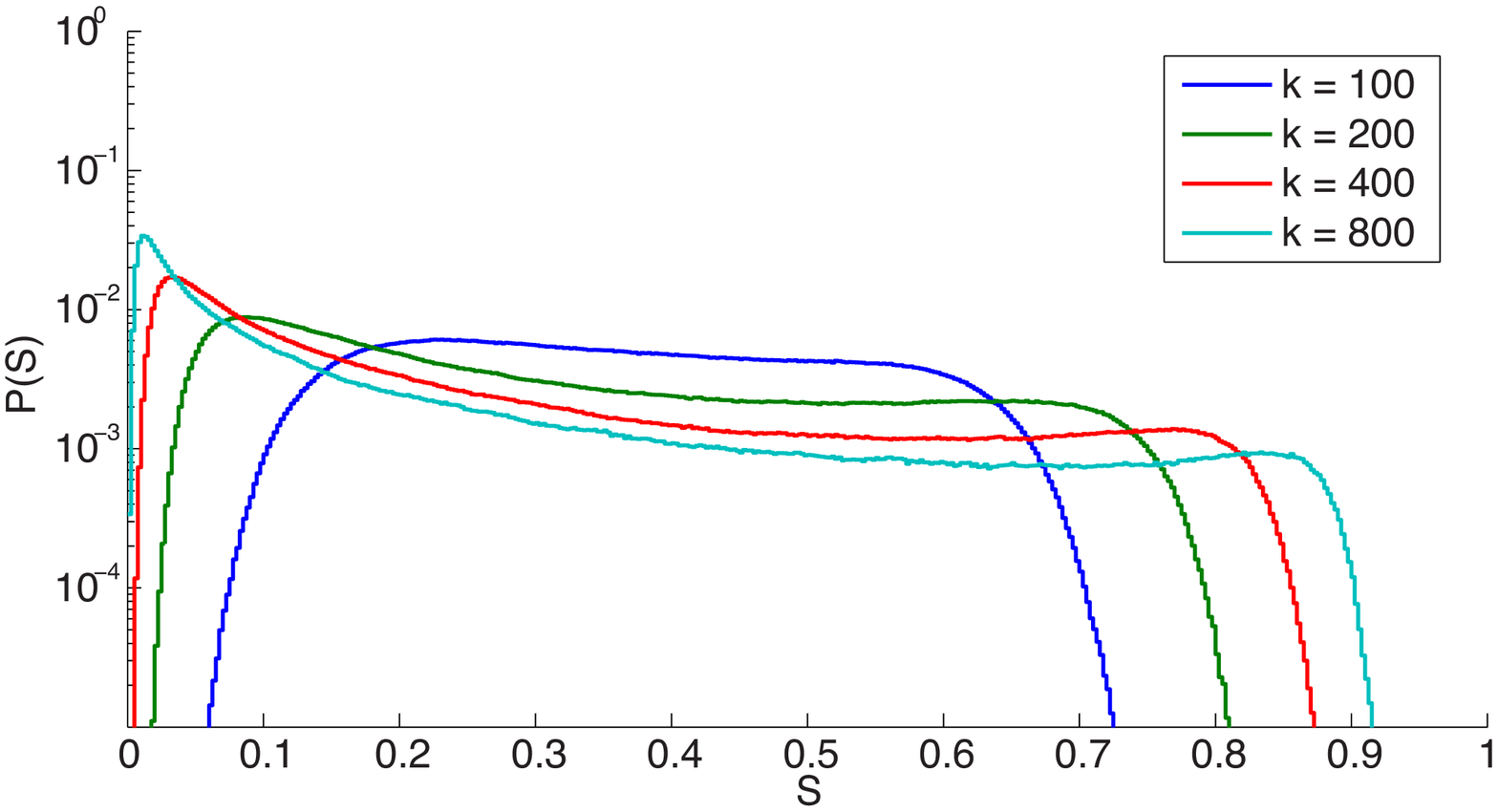}
	\caption{{\bf Form of $P(S)$}. (Left) While this diagrammatic form of $P(S)$ is all that is required for our analysis here, it compares well with the critical case in Fig. 1B of the main text. (Right) To further demonstrate that this form of P(S) is reasonable, empirically measured $P(S)$ distributions are shown for $N=10^{4}$, $\alpha=0.2$ and various values of $k$.}
	\label{fig-pofssketch}
\end{figure}

In order for activity to cease entirely, $S \sim \sqrt{\langle(S - \langle S \rangle )^{2} \rangle} \approx \sqrt{S/N}$ which implies that $S\approx 1/N$. While in this case the Fokker-Planck description is at the border of its range of validity, we presume that we can still use it for the purpose of obtaining a rough scaling estimate. Therefore, we write
\begin{equation}
	\text{(flux out)} \approx \text{(const.)} P(1/N) = \text{(const.)} P_{0} \exp{\left [- \frac{2N(\Lambda_{0} - 1)}{k(1-\alpha)} x\left(\frac{k}{N} (1-\alpha) \right) \right]}.
\end{equation}
Since we are primarily interested in scaling for large system size, we assume that $k / N \ll 1$, so $x\left[(1-\alpha)k/N\right] \approx x(0) = $ constant, and thus
\begin{equation}
	\text{(flux out)} \approx \text{(const.)} P_{0} \exp{\left[- \frac{2N(\Lambda_{0} - 1)}{k(1-\alpha)} x(0) \right]}.
\end{equation}
We now estimate $\tau = $ (total probability)/(probability flux out), and substitute the definition of $\Lambda_{0} -1 = \alpha / (1-2\alpha)$, yielding
\begin{equation}
	\tau = C_{1} \exp {\left[ C_{2} \left ( \frac{N \alpha}{k(1-\alpha)(1-2\alpha)} \right ) \right]}
\end{equation}
and we therefore argue that lifetime $\tau$ scales with a single scaling parameter $q = (N/k)\left[\alpha/(1-\alpha)(1-2\alpha)\right]$, as 
\begin{equation}
	\tau(q) = C_{1} \exp{[ C_{2} q ]}.
\end{equation}

\subsection*{Derivation of $\Lambda_{0}$.} 
To better understand the tendency for low activity to grow, we investigate $\Lambda_{0} = \lim_{S \to 0^{+}}\Lambda(S)$, which is presented as Eq.~(5) in the main text. This limiting case will consist of a single active node. A single active inhibitory node will produce zero additional active nodes, so the calculation of $\Lambda_0$ simplifies considerably. Thus we let $n_e = 1$ and $n_i=0$ in Eq.~(4), and substitute in the Poisson probability,  in which case the limit simplifies to  $\Lambda_{0} = \text{E}\left[\sigma(w)\right]  \langle k \rangle (1-\alpha),$ where $w$ is a draw from the link weight distribution. This expression has the following intuitive interpretation: the expected number of activated nodes immediately following a single active node is equal to the product of (i) the expected value of the transfer function after receiving a single excitatory input, (ii) the mean degree $\langle k \rangle$, and (iii) the probability that the single active node is excitatory. For the simple piecewise linear $\sigma$, and assuming $0\leq w\leq 1$, we get $\Lambda_{0} = \gamma\langle k \rangle (1-\alpha).$ Furthermore, since $\gamma \approx \lambda / [\langle k \rangle(1-2\alpha)]$, we conclude that
\begin{equation}
  \Lambda_{0} \approx \lambda\frac{1-\alpha}{1-2 \alpha}.
  \label{eq-lambda0}
\end{equation} 
Note that $\Lambda_0$ is only a function of the fraction of inhibitory nodes, and does not depend on the total number of nodes or mean degree. The factor $(1-\alpha)$ in the numerator of \eqref{eq-lambda0} reflects the fact that only excitatory nodes contribute to $\Lambda_0$. This may also be understood in terms of the excitatory-only subnetwork, as described in the next section.

\subsection*{Derivation of $\lambda^{+}$}

For large random networks, the largest eigenvalue can be well-approximated by the mean degree of the weighted network adjacency matrix \cite{restrepo}. Here we estimate the largest eigenvalue of a submatrix of a weighted network adjacency matrix. The full $N\times N$matrix, $A$, has mean degree $\langle k \rangle$, and its edge weights have magnitudes that are uniformly distributed with mean $\gamma$. A fraction $\alpha$ of columns of $A$, corresponding to the outgoing connections of inhibitory nodes, are negative. If $\gamma = \lambda / \left [ \langle k \rangle ( 1- 2\alpha) \right ]$ then the largest eigenvalue of $A$ will be approximately $\lambda$. We now calculate the largest eigenvalue $\lambda^{+}$ of the adjacency matrix that corresponds to only the connections among the excitatory population, $A^{+}$, again using the mean-degree approximation of Ref \cite{restrepo}:
\begin{align}
	\lambda^{+} & \approx \text{ mean weighted degree of } A^{+} \nonumber \\
	&= \text{ mean unweighted degree of } A^{+} \times \text{mean edge weight} \nonumber \\
	&= \langle k^{+}\rangle \gamma \nonumber \\
	&= (1-\alpha) \langle k \rangle \gamma \nonumber \\
	&= \lambda \frac{1-\alpha}{1-2\alpha}.
\end{align}
And combining this result with Eq.~\eqref{eq-lambda0}, we have $\Lambda_0 \approx \lambda^{+}$.

\subsection*{Figure S2 - Sample raster plot}

\begin{figure}[h]
	\includegraphics[width=1.0\linewidth]{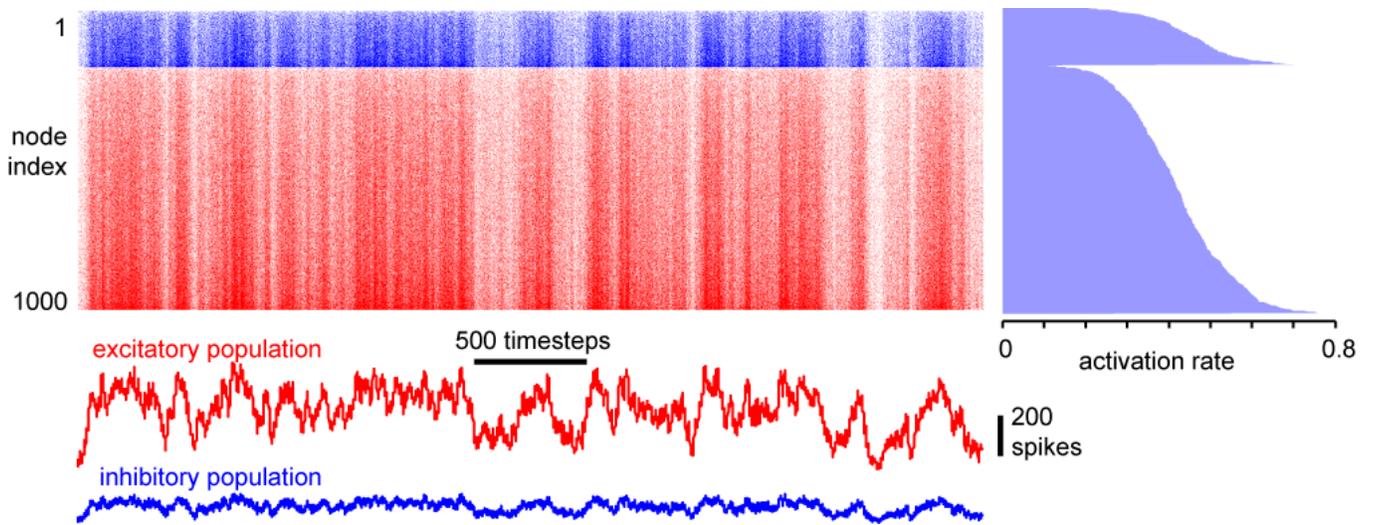}
	\caption{Sample dynamics with $N=1000$ nodes, $\langle k \rangle=50$, $\alpha=0.2$, and $\lambda=1$. The fluctuation of network activity over time can be captured by the raster plot whose traces over time (activation rate, right) and over nodes [time series of $S(t)$, bottom] may also be instructive. (top left) The raster plot shows a colored pixel when a node is active at a given time step and a white pixel otherwise. Rows are sorted first by inhibitory (blue) and excitatory (red) and then by firing rate. (top right) Most nodes are active at rates between 0.3 and 0.5. (bottom) Time series of the same dynamics, plotted separately for excitatory (red) and inhibitory (blue) populations.}
\end{figure}

\clearpage
\subsection*{Notes on the duration of thresholded critical branching process}

Critical branching processes are known to creat cascades with sizes distributed according to a power law with exponent $-3/2$ and with durations distributed according to a power law with exponent $-2$. In the main text, we make claims that our system operates in a critical regime, and show that for various inhibitory fractions $\alpha$ we get avalanches whose sizes are power-law distributed with the critical exponent. However, the durations do not appear to be power law distributed---even without proper statistical testing \cite{clauset} it is clear that the points do not fall on straight lines with slope of $-2$. We decided to investigate this further, with a short experiment: create time series from the simplest possible critical branching process, apply a threshold to define avalanches, and examine avalanche size and duration statistics. Data may be generated using very few lines of MATLAB code:
\begin{verbatim}
max_duration = 10000;                         % maximum avalanche duration
s_star = 128;                                 % threshold, S*
s = s_star;                                   % begin the process at the threshold
dura = 0;                                     % begin the process with 0 duration
size = s_star;                                % begin the process with s_star size
while (s>s_star && d < max_duration)          % while s is above the threshold
     s = binornd(2*s,1/2);	                   % binomial: 2*s trials with p=1/2
     dura = dura+1;                           % increment duration by 1
     size = size+s;                           % increment size by s
end
\end{verbatim}
Using this code, we generated $10^5$ avalanches with maximum duration of $10^5$ at each of the following values of $S^{*}$: 1, 2, 4, 8, 16, 32, 64, 128. Avalanche size and duration distributions are shown in Fig.~\ref{synth}. 

\begin{figure}[b]
	\centering
	\includegraphics[width=1.0\linewidth]{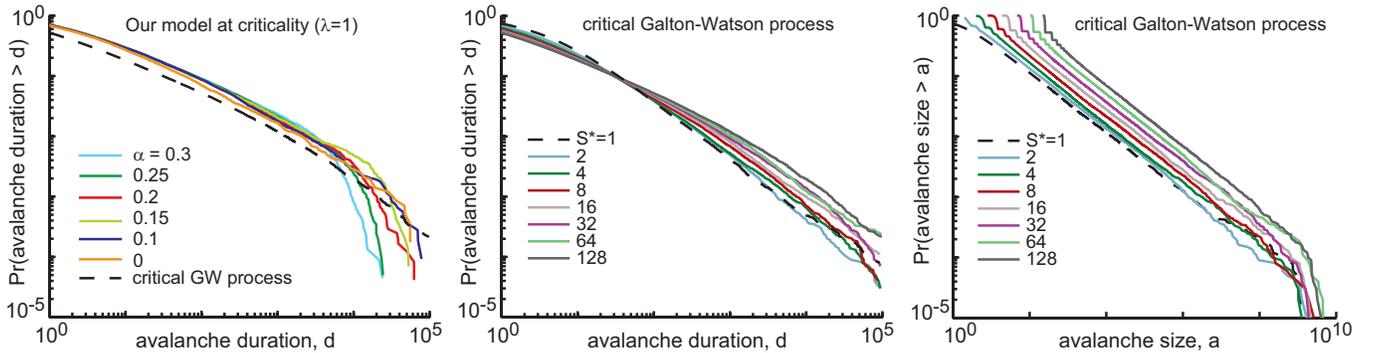}
	\caption{{\bf CCDFs of avalanche size and duration from thresholded critical branching processes}. (left) Avalanche durations from network data (colored lines) with the CCDF of the critical Galton-Watson branching process using $S^{*}=128$ (dashed line) overlaid. We did not statistically test the plausibility of these data being drawn from the same distribution, but show them here for visual comparison. (middle) Avalanche duration distributions deviate from power-law form when the threshold $S^{*}>1$. This implies that a power-law test of avalanche durations may not be a good test for criticality, as all data shown in both panels were generated from a critical branching process (see MATLAB code). (right) Avalanche size distributions for a variety of thresholds show the expected power-law form from critical branching theory, with power laws whose exponent is $-3/2$.}
	\label{synth}
\end{figure}

Duration distributions are not power-law distributed, except in the case where $S^{*}=1$, which corresponds to an unmodified critical branching process with the natural threshold of complete extinction. In this case, the exponent was $-1.96$ and statistical tests revealed that a power law is, statistically speaking, a plausible hypothesis for the data \cite{clauset}. However, the same statistical tests rejected the power-law hypothesis for duration distributions when $S^{*}>1$, leading us to conclude that critical branching avalanches do not have power-law distributed durations when a threshold is imposed. 

Size distributions, on the other hand, are power-law distributed, with exponents around $-3/2$, regardless of threshold. This is well-aligned with what is discussed and shown in the main text.  

Our first conclusion from this numerical experiment is that the avalanche size distribution is a much more reliable gauge of critical avalanches when avalanches are generated by fragmenting a continuous time series of readings into discrete events. 

Our second conclusion is that the observation that avalanche durations do not follow a $-2$ power law does not necessarily imply that the avalanches are not critical. Avalanche duration distribution is more complicated. Statistical tests exist for power laws, but since the exact form of the distribution of thresholded avalanches is not yet known, a statistical test for plausibility cannot be easily written down. This may have implications for many experimental measurements or empirical observations in which avalanches cease when they become unobservable or pass below an instrument's detection confidence limit.

\end{widetext}
\end{document}